\documentclass[12pt,manuscript]{emulateapj}
\usepackage{graphicx}
\usepackage{amssymb}
\usepackage{amsmath}
\usepackage{natbib}

\newcommand{\lnm}{\bf LN}

\shorttitle{Lognormal Distribution of Cosmic Voids}
\shortauthors{Pycke and Russell}

\begin{document}

\title{Lognormal Distribution of Cosmic Voids in Simulations and Mocks}


\author{E. Russell}
\affil{Division of Science and Mathematics, New York University Abu Dhabi, PO Box 129188, Abu Dhabi, UAE}
\email{er111@nyu.edu}

\and

\author{J-R Pycke}
\affil{Division of Science and Mathematics, New York University Abu Dhabi, PO Box 129188, Abu Dhabi, UAE}
\affil{Laboratoire de Math\'ematiques et Mod\'elisation d'\'Evry, Universit\'e d'\'Evry , PO Box 91037, \'Evry Cedex, France}
\email{jrp15@nyu.edu}

\begin{abstract}
Following up on previous studies, we here complete a full analysis of the void size distributions of the Cosmic Void Catalog (CVC) based on three different simulation and mock catalogs; dark matter, haloes and galaxies. Based on this analysis, we attempt to answer two questions: Is a $3$-parameter log-normal distribution a good candidate to satisfy the void size distributions obtained from different types of environments? Is there a direct relation between the shape parameters of the void size distribution and the environmental effects?
In an attempt to answer these questions, we here find that all void size distributions of these data samples satisfy the $3$-parameter log-normal distribution whether the environment is dominated by dark matter, haloes or galaxies. In addition, the shape parameters of the $3$-parameter log-normal void size distribution seem highly affected by environment, particularly existing substructures. Therefore, we show two quantitative relations given by linear equations between the skewness and the maximum tree depth, and variance of the void size distribution and the maximum tree depth directly from the simulated data. In addition to this, we find that the percentage of the voids with nonzero central density in the data sets has a critical importance. If the number of voids with nonzero central densities reaches $\geq \% 3.84$ in a simulation/mock sample, then a second population is observed in the void size distributions. This second population emerges as a second peak in the log-normal void size distribution at larger radius.
\end{abstract}

\section{Introduction}
The intricate formation of the large scale structure of the present-day Universe is formed by an interplay between random Gaussian fluctuations and gravitational instability. When gravitational instabilities start dominating the dynamical evolution of the matter content of the Universe, the structure formation evolves from a linear to a highly nonlinear regime. In this framework, voids are formed in minima and haloes are formed in maxima of the same primordial Gaussian field, and later on these features present different type dynamical evolutions due to their initial conditions in the nonlinear regime. This fact has been known since early studies that showed voids are integral features of the Universe \citep{chincarini,gregory,einasto,weygaert11}. \cite{sw,Esra1,Esra2} show that the distribution of voids can be affected by their environments. As a result, the void size distribution may play a crucial role to understand the dynamical processes affecting the structure formation of the Universe \citep{Croton2005,goldbergvogeley,Hoyle2005}.

The early statistical models of void probability functions (VPFs) \citep{fry86Voids,elizaldegaztanaga92} are based on the counts in randomly placed cells following the prescription of \cite{white79}. Apart from VPFs, the number density of voids is another key statistic to obtain the void distribution. Recently \cite{Pycke2016} show that void size distributions obtained from the Cosmic Void Catalog (CVC) satisfy a $3$-parameter log-normal probability function. This is particularly interesting, because observations and theoretical models based on numerical simulations of galaxy distributions \citep{Hamilton1985,colesjones91,Bernardeau92,Bouchet1993,Bernardeau94,Kofman,TaylorWatts2000,Kayo} show that the galaxy mass distribution satisfies a log-normal function rather than a Gaussian. Taking into account that voids are integral features of the Universe, it may be expected to obtain a similar distribution profile for voids. Apart from this, \cite{Pycke2016} discuss a possible quantitative relation between the shape parameters of the void size distribution and the environmental affects.

Following up on the study of \cite{Pycke2016}, we here extend their analysis of void size distributions to all simulated and mock samples of CVC of \cite{sutter}. The three main catalogs under study are dark matter, halo and galaxy catalogs. Therefore, we confirm that the system of $3$-parameter log-normal distribution obtained by \cite{Pycke2016} provides a fairly satisfactory model of the size distribution of voids. In addition to this, we obtain equations which satisfy linear relations between maximum tree depth and the shape parameters of the void size distribution as proposed by \cite{Pycke2016}.

\section{Void Catalog: Simulations and Mock Data}
Extending the study by \cite{Pycke2016}, we here fully investigate the void size distribution function statistically in simulations and mocks catalogs of the Public CVC of \cite{sutter}. It is useful to note that all the data of CVC are used here generated from a $\Lambda$ cold dark matter ($\Lambda$CDM) N-body simulation by using an adaptive treecode 2HOT \citep{sutter2014a,sutter2014b}. In addition, in all data sets voids are identified with the modified version of the parameter-free void finder ZOBOV \citep{Neyrinck2008,Lavaux12,sutter}. The data sets of CVC, we used here, can be categorized into three main groups;

\begin{itemize}
  \item Dark matter (DM) simulations are DM Full, DM Dense and DM Sparse. Although these dark matter simulations have the same cosmological parameters from the Wilkinson Microwave Anisotropy Probe (WMAP) seven year data release (DR$7$) \citep{komatsu2011} as well as the same snapshot at z=0, they have different tracer densities of $10^{-2}$, $4\times10^{-3}$, and $3\times10^{-4}$ particles per $(Mpc/h)^3$ which are respectively DM Full, DM Dense, and DM Sparse. Also the minimum effective void radii $R_{eff,min}=5,  7$ and $14$ Mpc/h are obtained from the simulations for DM Full, DM Dense and DM Sparse respectively.

  \item Halo catalog in which two halo populations are generated; Haloes Dense and Haloes Sparse. In the halo catalog the halo positions are used as tracers to find voids. The minimum resolvable halo mass of Haloes Dense is $1.47\times 10^{12}$ $M_{\odot}/h$ while the minimum resolvable halo mass of Haloes Sparse data set is $1.2\times 10^{13}$ $M_{\odot}/h$. In addition, the minimum effective void radii of Haloes Dense and Sparse are $R_{eff,min}=7, 14$ Mpc/h respectively. The main reason to construct these halo populations with different minimum resolvable halo masses is to compare the voids in halos to voids in relatively dense galaxy environments, see \cite{sutter2014a} for more details.

  \item Galaxy catalogues; there are two galaxy mock catalogues which are produced from the above halo catalog by using the Halo Occupation Distribution (HOD) code of \cite{Tinker2006} and the HOD model by \cite{Zheng07}. These galaxy mock catalogs are called HOD Dense and HOD Sparse \citep{sutter2014a}. The HOD Dense catalog has $9503$ voids with effective minimum radii $R_{eff,min}= 7$ Mpc/h and includes relatively high-resolution galaxy samples with density $4\times10^{-3}$ dark matter particles per cubic Mpc/h matching the Sloan Digital Sky Survey (SDSS) DR$7$ main sample \citep{Strauss2002} using one set of parameters found by \cite{Zehavi2011} ($\sigma_{log M}=0.21$, $M_{0}=6.7\times10^{11} h^{-1}M_{\odot}$, $M^{\prime}_{1}=2.8\times10^{13}h^{-1}M_{\odot}$, $\alpha=1.12$). The HOD Sparse mock catalog consists of $1422$ voids with $14$ $Mpc/h$ effective minimum radii ($R_{eff,min}= 14$ $Mpc/h$) and this void catalog represents a relatively low resolution galaxy sample with density $3\times 10^{-4}$ particles per cubic $Mpc/h$ matching the number density and clustering of the SDSS DR$9$ galaxy sample \citep{Dawson2013} using the parameters found by \cite{Manera2013} ($\sigma_{\log M}=0.596$, $M_{0}=1.2\times10^{13}$ $h^{-1}M_{\odot}$, $M^{\prime}_{1}=10^{14} h^{-1}M_{\odot}$, $\alpha=1.0127$, and $M_{min}$ chosen to fit the mean number density). In addition to this another mock galaxy catalog is used here; N-body Mock catalog which is a single HOD Mock galaxy catalog in real space at $z=0.53$, generated by a dark matter simulation of $4096^3$ particles (with a particle mass resolution $7.36\times 10^{10}$ $h^{-1}M_{\odot})$ in a $4$ Gpc/h box and is tuned to SDSS DR$9$ in full cubic volume by using the HOD parameters found in \cite{Manera2013} and it consists of $155,196$ voids \citep{sutter2014b}. Although the N-body Mock catalog is processed slightly differently than HOD Sparse and HOD Dense, it is a HOD mock catalog and it uses Planck first- year cosmological parameters \citep{PlanckCollaboration}.
  \end{itemize}
In the following section, we examine the above data sets from a statistical perspective such as histograms, parameters of location (range, mean, median), mode or dispersion (standard deviation) and shape (skewness, kurtosis) by following the previous study of \cite{Pycke2016}. From a statistical perspective, we also investigate the connection between the distribution and the environment of void populations.

\section{Statistical Properties of Void Distributions}

As a first step, the raw data plots of void size distributions are obtained for DM Full, DM Dense, DM Sparse, Haloes Dense and Haloes Sparse. Note that the void size distributions for HOD Dense, HOD Sparse and N-body mock data sets are discussed in great detail in \cite{Pycke2016}. In the raw void size distributions, an unexpected local peak is observed around the value $20$ Mpc/h in the DM Full sample and in the DM Dense sample around $27$ Mpc/h, see Fig 1 Upper Left and Lower Left Panels. A similar behavior is observed by \cite{Pycke2016} in the N-body Mock sample around the value $50$ Mpc/h.

\begin{figure*}
\centering
\begin{tabular}{ll}
\includegraphics[scale=0.55]{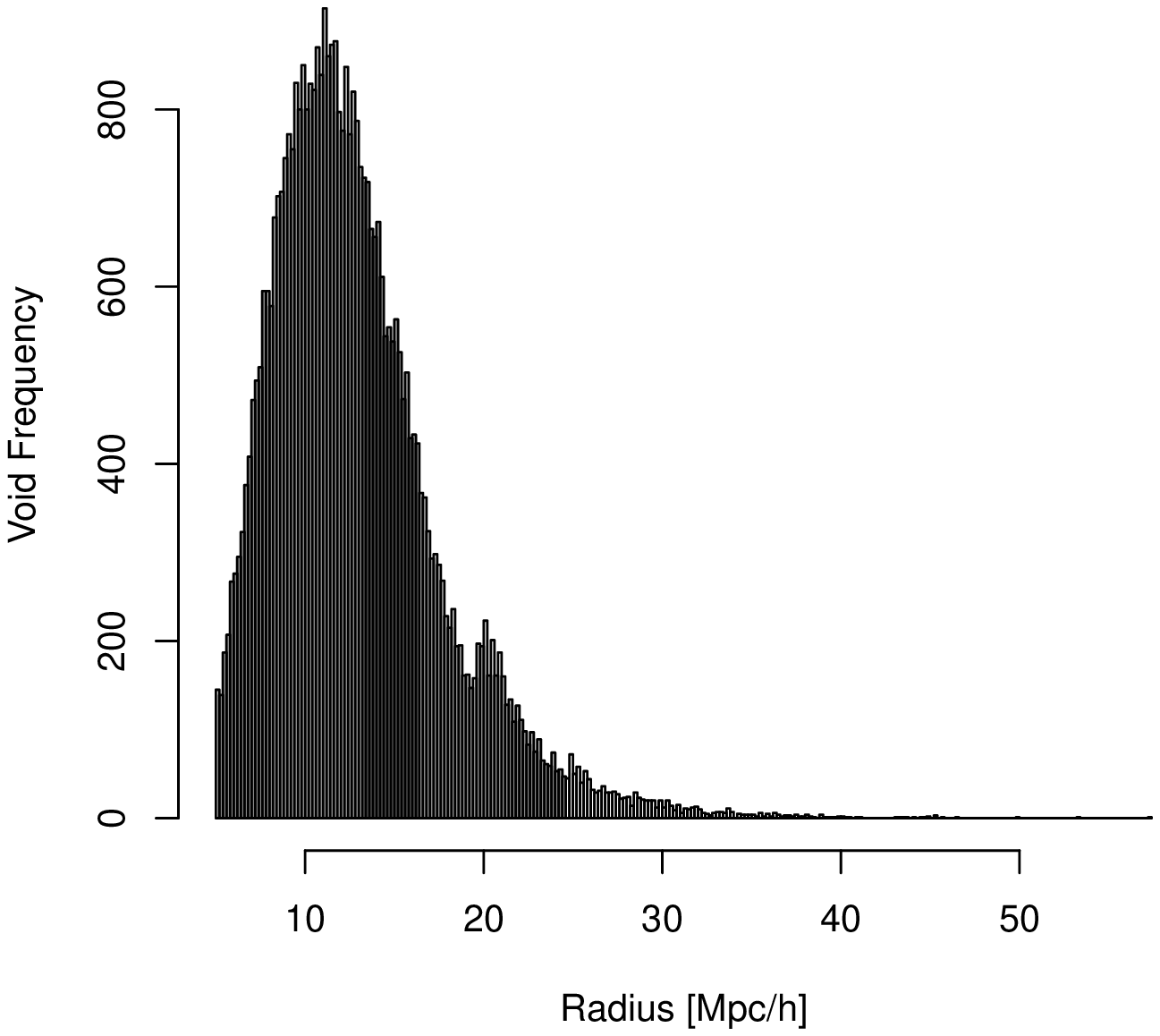}
\includegraphics[scale=0.55]{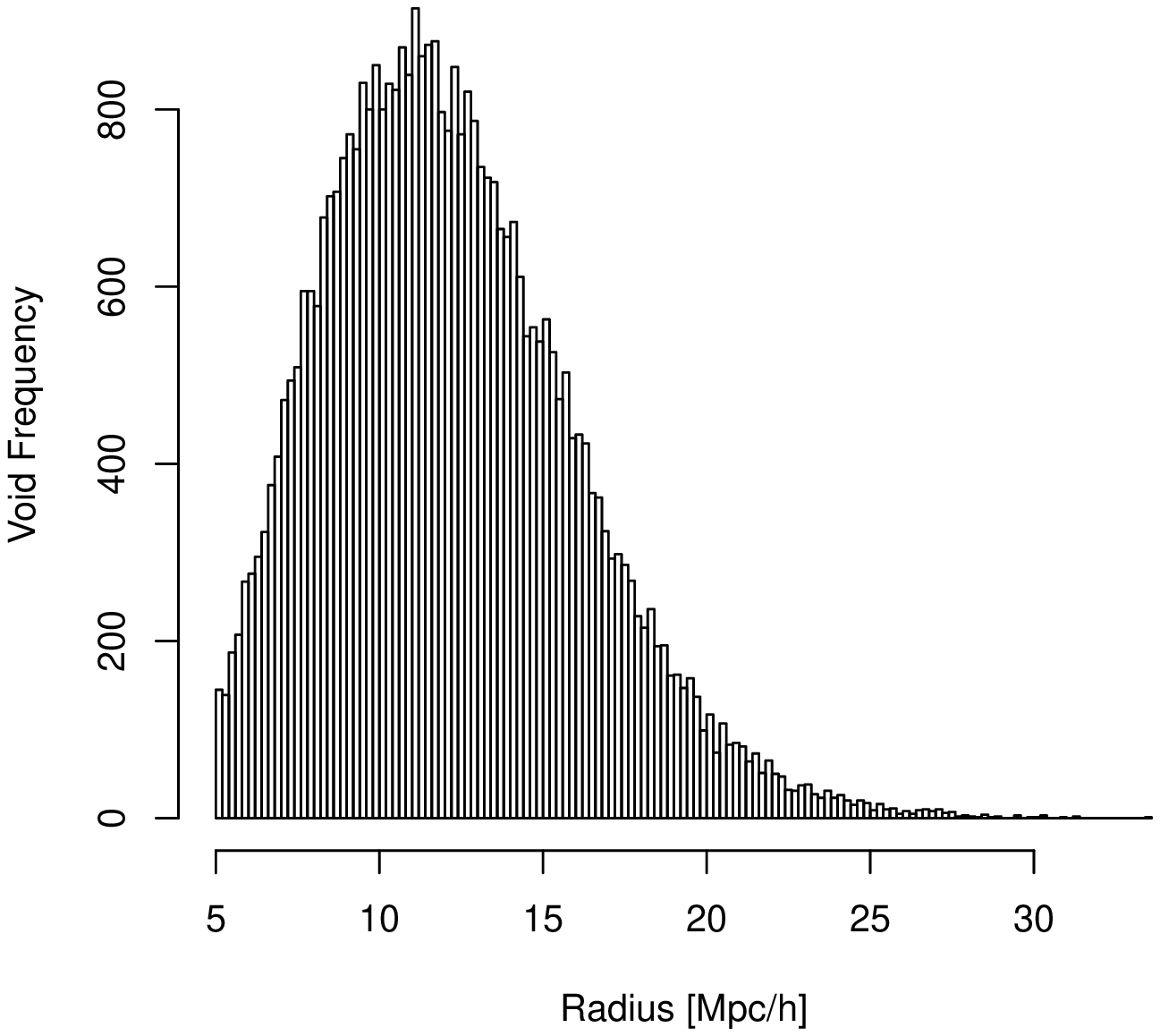}
\\
\includegraphics[scale=0.55]{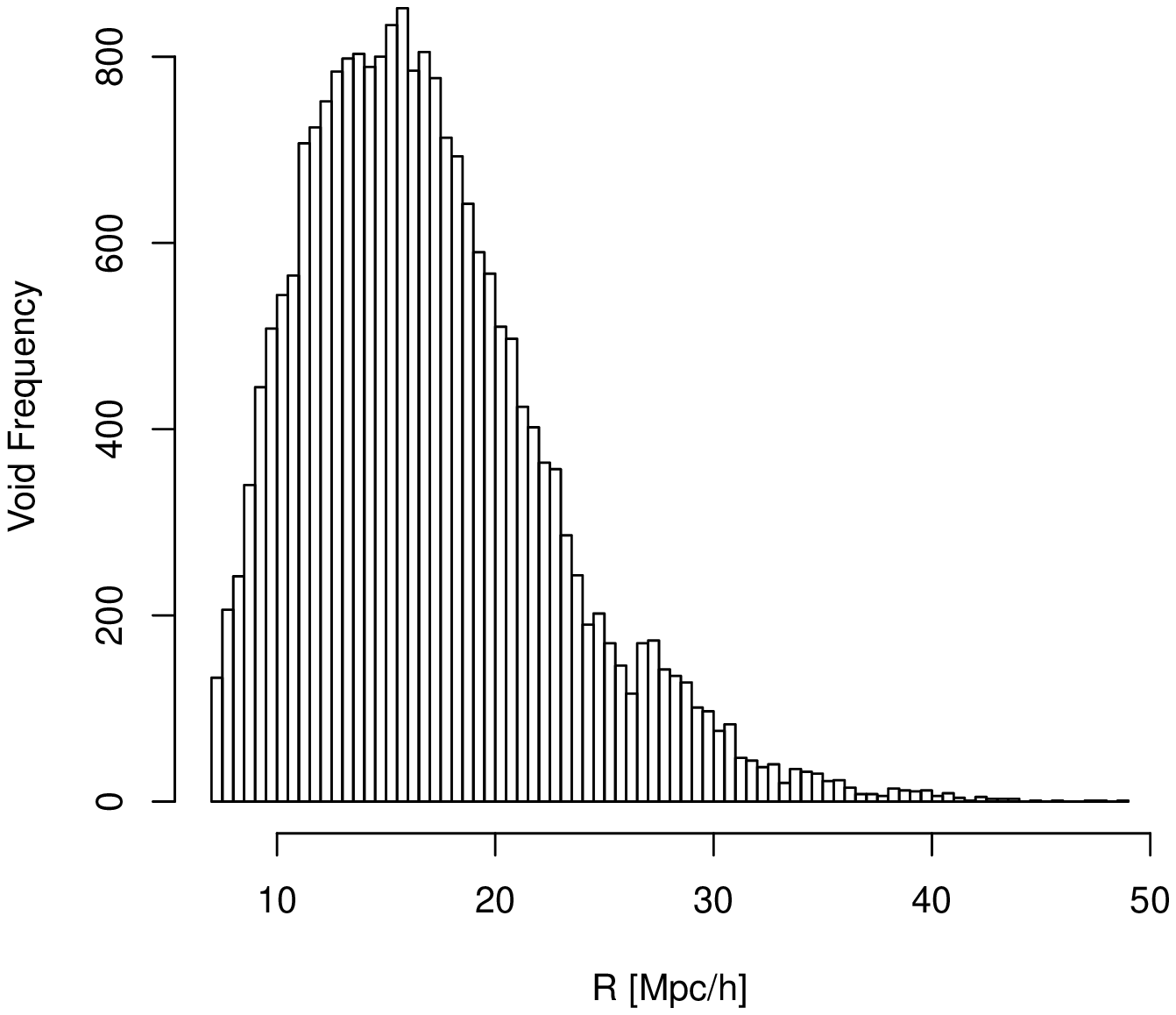}
\includegraphics[scale=0.55]{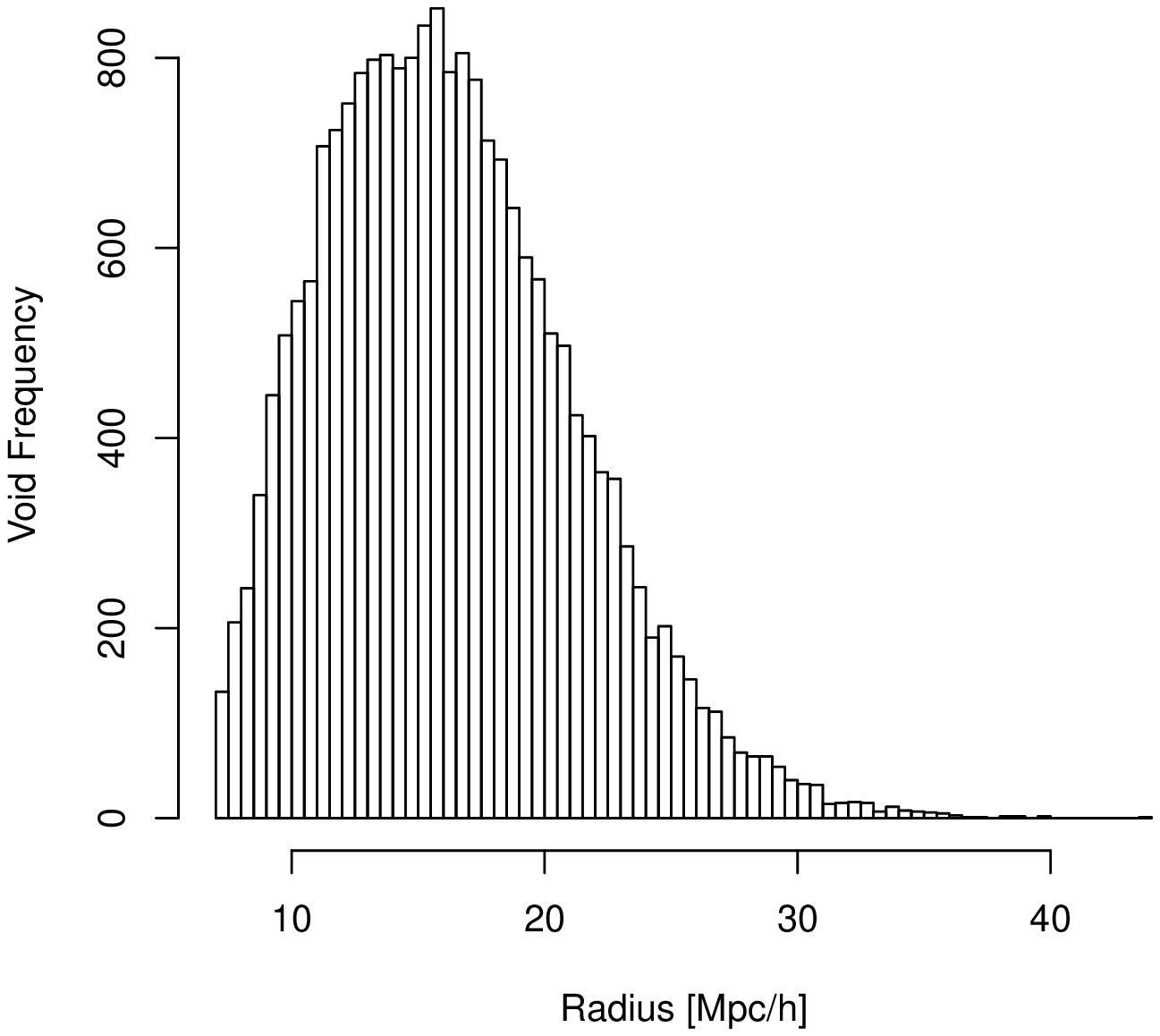}
\end{tabular}
\caption{Void size distributions of DM Full (Upper Panels) and DM Dense (Lower Panels) data sets. While Upper and Lower Left panels present the peak formations due to the presence of the second void population, the Right panels show their sub-samples respectively consisting of the void populations with only zero central density.}
\end{figure*}
It is crucial to mention that the samples, DM Full and DM Dense, are high resolution data sets compare to the samples we investigate here; DM Sparse, Haloes Dense, Haloes Sparse. We find that the low resolution samples; DM Sparse, Haloes Dense show single populations (see Fig 2).

\begin{figure*}
\centering
\begin{tabular}{ll}
\includegraphics[scale=0.6]{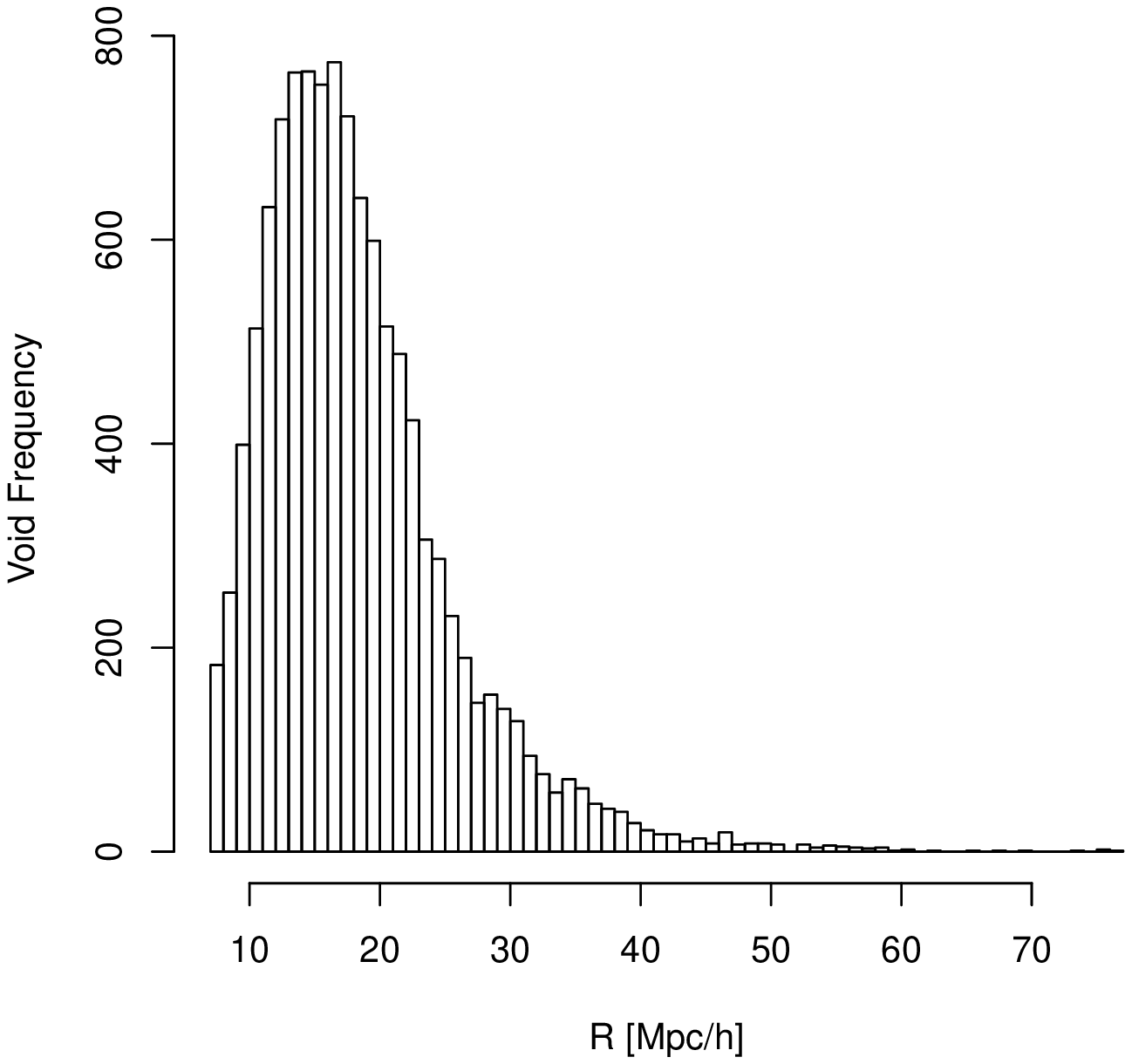}
\includegraphics[scale=0.6]{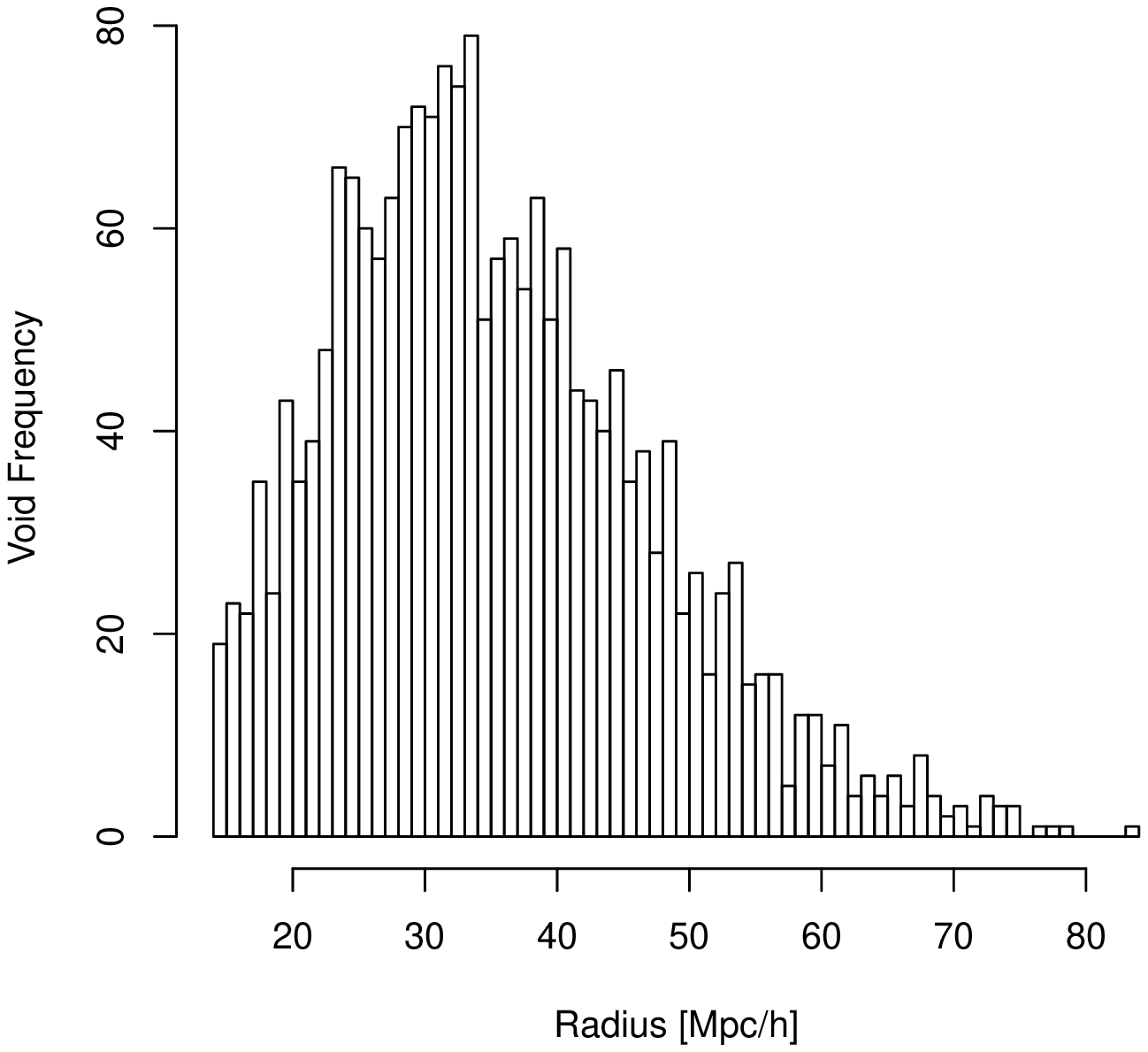}\\
\includegraphics[scale=0.6]{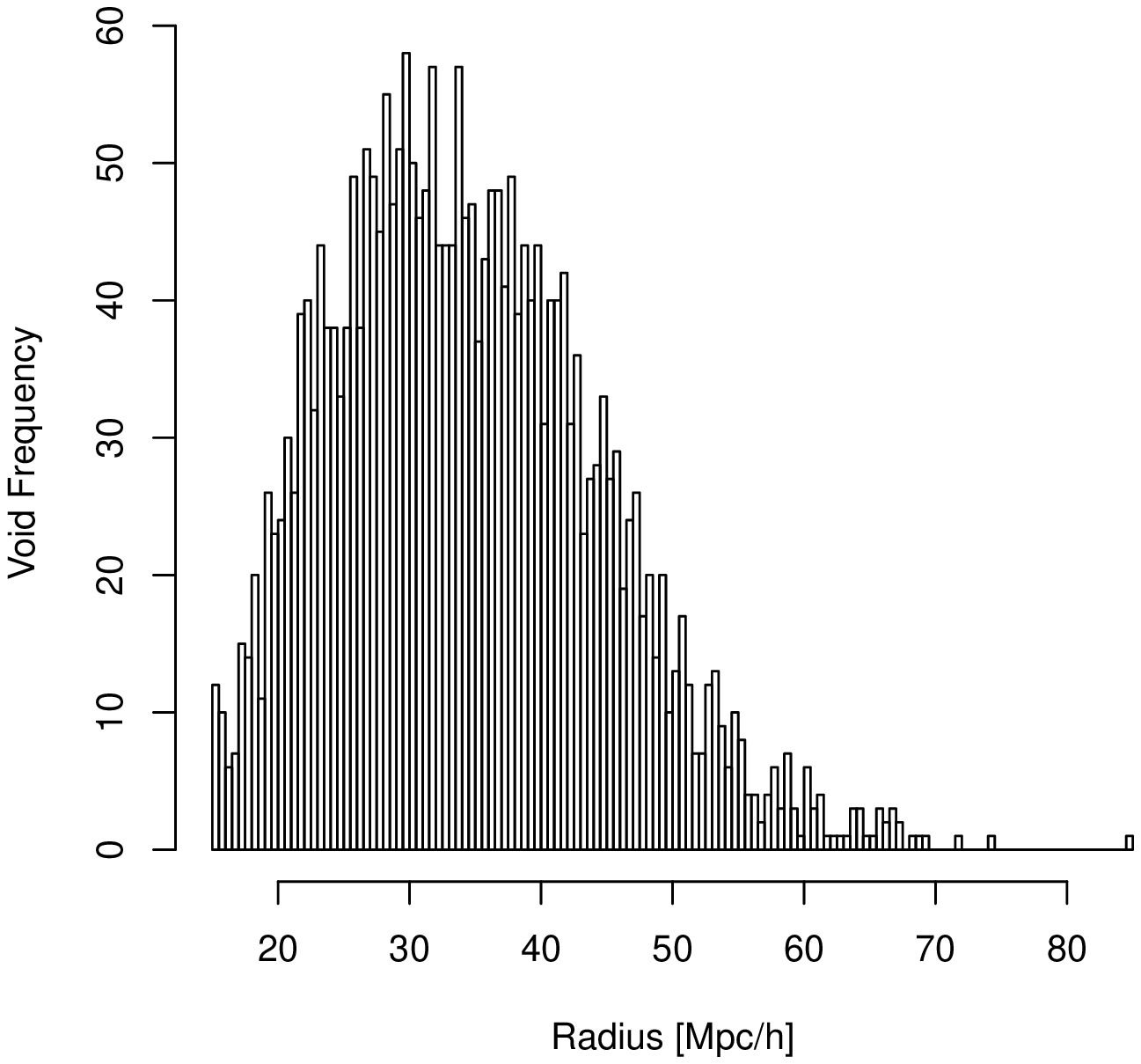}
\end{tabular}
\caption{Void size distributions of Haloes Dense, Haloes Sparse and DM Sparse from Upper Right to Lower Left, respectively.}
\end{figure*}
In the previous study, \cite{Pycke2016} show that HOD Sparse and HOD Dense samples also present single population void size distributions. Note that our goal is not to compare the invariant distributions in Fig 2, but to find the reason of the emergence of the second population (Fig 1) versus single population in the void size distributions of CVC. After analyzing the components of the data sets, we realize that the second peak occurs due to a second void population in the samples. We find that in both cases, the second void population has a higher central density than the dominant voids of the distributions which have zero central densities. Following the same strategy of \cite{Pycke2016} to investigate these samples in terms of statistical properties, the second peaks in DM Full and DM Dense are excluded by excluding the voids with nonzero central densities from the data sets. Following this procedure, the new sub samples of DM Full and DM Dense are obtained. As is seen in Fig 1 Upper and Lower Right Panels, the data samples present void size distributions obtained by single void populations. Although we eliminate the second peak to investigate these data sets in a statistical frame work, there are two questions left to answer to understand the occurrence of more than one void population in the void size distributions; Why cannot we observe a second peak in the rest of the data sets, DM Sparse, Haloes Dense, Haloes Sparse (see, Fig 2) although all samples include some fraction of voids with nonzero central densities? Is it possible to find a criterion of observing two different void populations in the void size distributions due to their central densities, particularly in the simulation and mock data?

Here we attempt to address these questions. Taking into account that N-body Mock, DM Full and DM Dense are high resolution data sets, it is expected to observe more substructures in the simulations, which leads to the formation of two different void populations particularly in terms of central densities indicating two different void environments. We find that emerging of multiple populations in the void size distributions is directly correlated with the number of voids with nonzero central densities, especially with the highest cental densities, $\rho_{cent} \in[0.2, 0.09]$. Table 1 presents the total number of voids of the samples and the percentage of voids with nonzero central densities in the range of [0.2-0.09] which includes the densest voids in each sample.

\begin{table*}
\caption{Percentages of number of voids with non zero central densities $\rho_{cent}$ in the interval $[0.2,0.09]$ and the total number of voids, $N_{Tot-voids}$, in each sample.}
$$\begin{array}{|c|c|c|}\text{Sample}& \text{Percentages}\phantom{a}\text{of}\phantom{a}N_{\text{voids}}\phantom{a}(\rho_{\text{cent}}\in [0.2,0.09])& N_{\text{Tot-voids}}\\\hline
\text{HOD Dense}&2.66 & 9,503\\
\text{HOD Sparse}&2.65 &1,422\\
\text{DM Full}& 5.32 & 42,948\\
\text{DM Dense}&3.84& 21,865 \\
\text{DM Sparse}&\approx 0.54 & 2,611 \\
\text{Haloes Dense}&3.03 &11,384\\
\text{Haloes Sparse}&2.65 & 2,073\\
\text{N-body Mock}&\approx5 & 155,196\\
\hline
\end{array}
$$
\end{table*}
As is seen from Table 1 the contribution of the denser central densities are low compared to voids with zero central densities $ < 3.84\%$ in all samples that present a single population. What we observe is that the second population forms a peak in the void size distribution as long as the percentage of voids with nonzero density $\geq 3.84\%$ in the overall data set (see Table 1). It seems that the contribution of voids with dense central densities to the overall void population holds an answer to the questions of the formation of the second or even more populations occurring in the void size distributions.

After reducing the DM Full and DM Dense samples to a single population, we can now investigate the statistical properties of the void size distributions;

\begin{itemize}
  \item Mean ($\overline{r}$):
  \begin{eqnarray}
  \overline{r}=\frac{1}{N}\sum_{i=1}^{N} r_i
    \end{eqnarray}
    where N is the total number of voids and $r_i$ the radius/size of each void in a given sample.
  \item Centered Moments:
  \begin{eqnarray}
  m_k=\frac{1}{N}\sum_{i=1}^N(r_i-\overline  r)^k, k=2,3,...
    \end{eqnarray}
    For instance $m_{3}$ and $m_{4}$ are related to skewness and kurtosis, respectively but they are influenced by the unit of measure.
  \item Variance $m_2$; measurement of the dispersion of the data. Using variance and the higher moments $m_{3}$ and $m_{4}$, the skewness and kurtosis are formulated,
  \begin{itemize}
    \item skewness $b_{1}$; measurement of the degree to which a distribution is asymmetrical.
    \begin{eqnarray}
    b_1=\frac{m_3^2}{m_2^3}
    \end{eqnarray}
    \item kurtosis $b_2$ is a measurement of the  degree of the peakedness.
    \begin{eqnarray}
      b_2=\frac{m_4}{m_2^2}
      \end{eqnarray}
      \end{itemize}
\end{itemize}
(see in \cite{kotz} formulas $(1.235)-(1.236)$ p. 51 and in \cite{stuart1} $(3.85)-(3.86)$ p. 85). The shape parameters of the void size distributions for each sample are presented in Table 2.

\begin{table*}
\caption{First moments, skewness $b_1$, kurtosis $b_2$ and maximum tree depth (from \cite{sutter2014a}) of the sample distributions. Note that the moments of the N-body Mock, DM Full and DM Dense distributions are computed for a single population.}
$$\begin{array}{|c|c|c|c|c|}\text{Sample}&\text{HOD Sparse}& \text{HOD Dense} &\text{N-body Mock}&\text{Haloes Dense}\\
\hline
\overline{r}&40.4&16.7&32.0&18.3\\
m_2&236 &40.1 &96.0&52.5\\
m_3&4,420 &386 &899&567\\
m_4&307,000 &11,100 &38,100&19,400\\
b_1&1.49 &2.32&0.912&2.23\\
b_2&5.52 & 6.93 &4.13&7.07\\
\text{Merging\phantom{a}Tree\phantom{a}Depth}& 4 & 10 &-&7\\
\hline
\text{Sample}&\text{DM Dense}&\text{DM Full}&\text{Haloes Sparse}&\text{DM Sparse}\\\hline
\overline{r}&16.3&12.2& 36.4& 34.4\\
m_2&24.5&24.5&178&102\\
m_3&73.3&32.5&2,190&596\\
m_4&1,970&736&127,000 & 34,600\\
b_1&0.365&0.434&0.841&0.328\\
b_2&3.29&3.49&3.96&3.28\\
\text{Merging\phantom{a}Tree\phantom{a}Depth}& 3 &-& 2 & 0\\
\hline
\end{array}
$$
\end{table*}
According to Table 2, all the void size distributions share the property of being significantly positively skewed ($m_3>0$) while the values of the kurtosis indicate a leptokurtic ($b_2>3$) behavior. Note that a leptokurtic type distribution is characterized by a high degree of peakedness (see \cite{sheskin} p. 16-30 for more details). The same result of positively skewed and leptokurtic void size distributions is also pointed out by \cite{Pycke2016} for three samples of CVC.

We would like to extend and modify here the discussion from \cite{Pycke2016} in which they propose the maximum tree depth as an environmental indicator. Note that \cite{sutter2014a} define the maximum tree depth as the length from root to tip of the tallest tree in the hierarchy, and it indicates the amount of substructures in the most complex void in the sample. Taking one step further from the point of \cite{Pycke2016} and using the maximum tree depth values from \cite{sutter2014a}, we here provide a simple linear relation indicating that the shape of the void size distribution is strongly correlated with the maximum tree depth of the simulated void catalogs. Table 2 also presents the maximum tree depth as well as the void size distribution parameters of the samples; HOD Dense, HOD Sparse, Haloes Dense, Haloes Sparse, DM Dense and DM Sparse. As is seen, in our computations the samples, N-body Mock and DM Full are not taken into account since the  shape parameters of their void size distributions are obtained after excluding the second populations in the samples in order to provide the correlations without any interference. Based on the maximum tree depth and the skewness parameters of the samples from Table 2 we obtain a linear function,

\begin{eqnarray}
b_{1}&=&0.210408(\pm0.03236 )MTD +0.473522(\pm 0.1881 ),\nonumber\\
R^2&=&0.9337,
\label{eqn:MTD-skew}
\end{eqnarray}
here MTD refers to the maximum tree depth and $R^2=0.9337$ is the regression of the data. According to equation \ref{eqn:MTD-skew}, when the amount of substructures is high (high MTD value), the void size distributions tend to be more positively skewed. Also, if one obtains the maximum tree depth of a sample, then it is possible to have a fairly good estimation of the amount of skewness of the void size distribution. As is seen in Fig 3, the higher the value of MTD, the more skewed the distribution is. This may lead to the fact that the skewness of a void size distribution can be a good indicator of the amount of substructures in a sample.

\begin{figure*}
\centering
\includegraphics[scale=0.8]{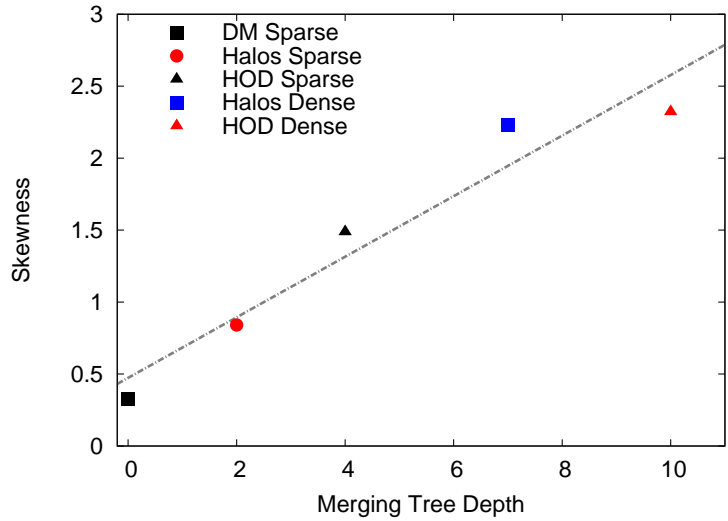}
\caption{Relation between skewness and maximum tree depth for DM Sparse, Haloes Dense, Haloes Sparse and HOD Sparse and HOD Dense.}
\end{figure*}
On the other hand, we must take into account a crucial parameter that can affect skewness-MTD relation. This parameter is the minimum radius-cut in the data sets. Basically there two main density-based criteria which are imposed at different stages of the data production of CVC; the first threshold cut comes from ZOBOV in which voids only include as members Voronoi cells with
density $ < 0.2$  the  mean  particle  density \citep{sutter2014a}. In the second density criterion
voids with mean central densities $< 0.2$ the mean particle density are included only. This is particularly an important criterion since it gives an insight about the radius cut. \cite{sutter2014a} gives the central density
within a sphere with radius,

\begin{eqnarray}
R=\frac{1}{4}R_{eff},
\label{eqn:radiuscut}
\end{eqnarray}
\noindent
in which the effective radius $R_{eff}$ is obtained by the following equation \citep{sutter2014a},

\begin{eqnarray}
R_{eff}=\left(\frac{3}{4\pi}\right)^{1/3}.
\label{eqn:radiuscuteff}
\end{eqnarray}
\noindent
Here V stands for total void volume. \cite{sutter2014a} points out that they ignore voids with
$R_{eff}$ below the mean particle spacing of the tracer population. As a result, this minimum radius cut constraint by the density criteria can affect the skewness-MTD relation.
\noindent
Apart from the skewness-MTD linear relation, we investigate the correlation between variance/dispersion and MTD. Hence, It is found that the correlation between MTD and variance shows a distinction between Sparse and Dense samples; see Fig 4. Because sparse data show high dispersion by its nature, converse to the dense data, it is an expected result to observe two main dispersions. As is seen in Fig 4, while sparse data show a fairly good (with high regression, $R^2=0.9938$) linear relation between MTD and variance (black line Fig 4),

\begin{eqnarray}
var(m_{2})_{Sparse}&=&33.438(\pm 2.634) MTD +105.041(\pm 6.801),\nonumber\\
R^2&=&0.9938,
\label{eqn:MTD-variance-sparse}
\end{eqnarray}
the variance of the dense data sets do not give enough information about the relation between MTD and dispersion (red line in Fig 4) although fitting to the dense data points give a linear line without an error.

\begin{eqnarray}
var(m_{2})_{Dense}=-4.134 MTD +81.392,\phantom{a} R^2=1.
\label{eqn:MTD-variance-dense}
\end{eqnarray}
This result is mathematically inconsistent since the variance of a distribution is always a positive number by its definition. That is why variance cannot accept negative values as it is stated by equation (\ref{eqn:MTD-variance-dense}). It seems that we need more data points to show a direct relation between MTD and the variance. Therefore we here cannot conclude that there is a direct relation between the variance and the MTD of the dense data sets.

\begin{figure*}
\centering
\includegraphics[scale=0.8]{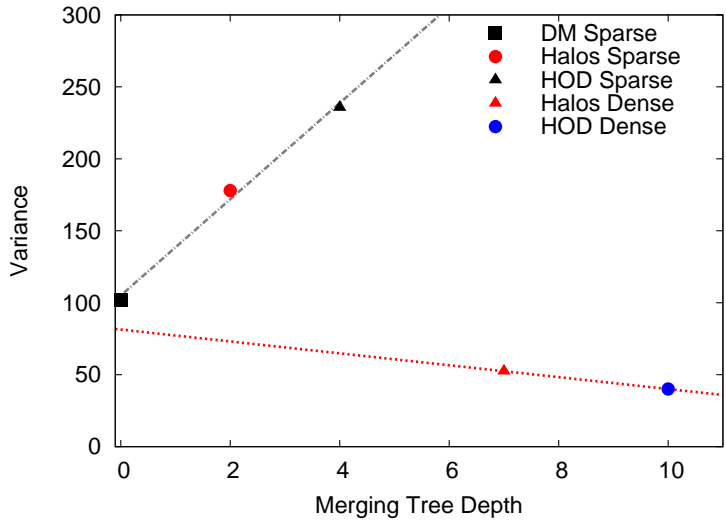}
\caption{Relation between MTD and variance of the sparse and dense data sets.}
\end{figure*}
We also extend the data points in the skewness and kurtosis plane $(b_1, b_2)$ given by \cite{Pycke2016} shown as the log-normal line observed in Fig 5. Fig 5 shows the void size distributions generated from the simulations and mock samples of CVC and these distributions can be considered to behave as log-normal distributions with respect to their skewness and kurtosis.

\begin{figure*}
\centering
\includegraphics[scale=0.9]{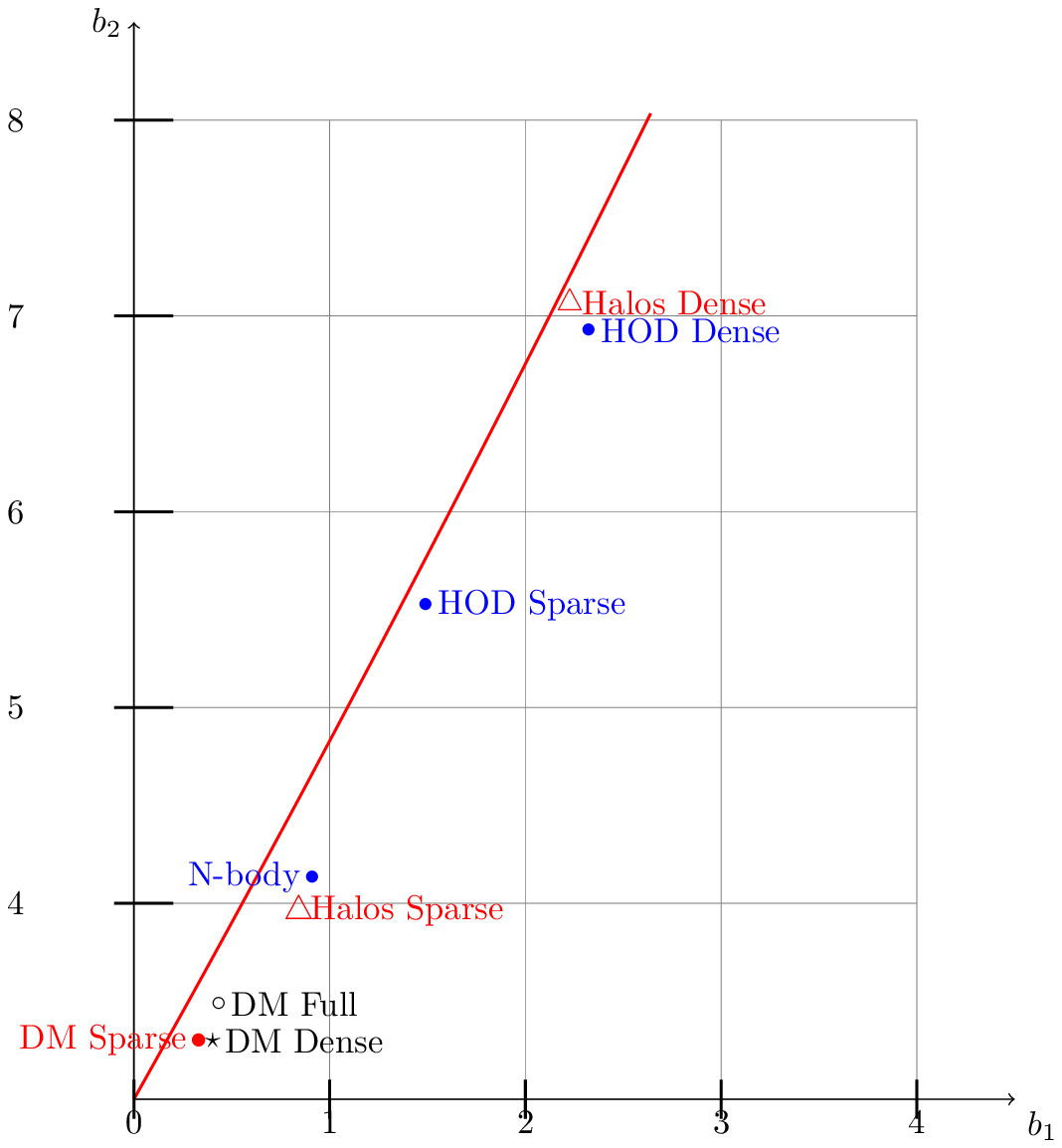}
\vspace{-120mm}
\caption{Kurtosis $b_{2}$ versus skewness $b_{1}$ parameters of the sample data in which solid red line represents the standard $3$-parameter log-normal distribution. Note that here the parameters of the N-body \citep{Pycke2016}, DM Full and DM Dense samples are obtained after excluding the second populations to obtain statistical properties.}
\end{figure*}
Extending the data sets from the previous paper of \cite{Pycke2016}, we confirm here that the $3$-parameter log-normal distribution with random variable $\mathbf R=\lnm(\theta,\zeta,\sigma)$,

\begin{equation}
p_{\theta,\zeta,\sigma}(R)=\frac{e^{-\frac{\left\{\log(R-\theta)-\zeta\right\}^2}{2\sigma^2}}}{(R-\theta)\sigma\sqrt{2\pi}},\quad R>\theta,
\label{eqn:void-distribution}
\end{equation}
fits the void size distributions obtained from the simulated and mock data sets of CVC. It is important to note that
in this study, following up on previous paper by \cite{Pycke2016}, we use the moment method to obtain the fit. This method is one of the standard methods in the field of estimation. We prefer this method to the maximum likelihood method due to the technical uncertainties and difficulties related to the maximum likelihood method by following \cite{kotz}, see page 228. In addition to this, we here do not provide the standard goodness of fit tests, such as Anderson-Darling, Cram\'er-von Mises or Kolmogorov to the data samples of CVC for the following reasons. \cite{sheskin} cites the study of \cite{conover80,conover99} which point out that if one employs a large enough sample size, almost any goodness-of-fit test will result in rejection of the null hypothesis. \cite{conover80,conover99} also states that if the sample data are reasonably close to the hypothesized distribution, one can probably operate on the assumption that the sample data provide an adequate fit for the hypothesized distribution. Taking into account the size of the simulated and mock samples as well as the uncertainties of the construction of CVC (for example, minimum-radius cuts), the moment method provides a straightforward tool to obtain the parameters of the distribution (see the formulas of the first three sample moments in \cite{kotz}, page 228).

In above distribution formula \ref{eqn:void-distribution}, a random variable ${\mathbf R}$ is defined by $\lnm(\theta,\zeta,\sigma)$ if $\log(\mathbf R-\theta)$ follows a Gaussian distribution with mean $\zeta$ and variance $\sigma^2$ given by \cite{kotz}. \cite{kotz} describe the characteristics of a random variable $\lnm(\theta,\zeta,\sigma)$ as,

\begin{enumerate}
  \item range: $(\theta,\infty)$,
  \item mode: $\theta+e^{\zeta-\sigma^2}$,
  \item median: $\theta+e^{\zeta}$,
  \item mean: $\theta+e^{\zeta+\sigma^2/2}$.
\end{enumerate}
Some characteristics of a log-normal random variable as well the relations between the shape parameters and estimators to fit the data samples are discussed in great detail by \cite{kotz}. Using the above characteristics with the shape parameters as indicated by \cite{kotz}, one can obtain the estimators. We here also provide the estimators of the log-normal void size distributions of Haloes Dense, DM Dense and DM Full data sets in Table 3.

\begin{table*}
\caption{Estimates of the $3$-parameter log-normal distributions for Haloes Dense, DM Dense and DM Full samples.}
$$\begin{array}{|c|c|c|c|c|c|c|}Sample&Haloes Dense &DM Dense &DM Full\phantom{a}(single\phantom{a}population)\\
\hline
\hat \theta&2.67&-8.61&-5.43\\
\hat \zeta&2.65&3.20&2.85\\
\hat \sigma&0.441&0.196&0.213\\
\hline
\end{array}
$$
\end{table*}
The estimates of the $3$-parameter log-normal void size distributions of HOD Dense, HOD Sparse and N-body Full are given by \cite{Pycke2016}.  As aforementioned, Haloes Sparse and DM Sparse samples are not taken into account in the further analysis due to their highly fluctuating distributions, see Fig 2. As is seen in Table 3, $\theta$ parameters of DM Dense and DM Full samples accept negative values. On the other hand, these negative $\theta$ values do not cause any inconsistency in the distributions as long as $R> \theta$. Therefore $\log(\mathbf R-\theta)$ is always defined for the samples even with negative $\theta$ values.

It seems that the $3$-parameter log-normal distribution is a natural candidate to fit the size distributions of the void samples. \cite{Pycke2016} already show that HOD Dense, HOD Sparse and N-body Mock also satisfy this $3$-parameter log-normal void size distribution. The goodness-of-fit of our model is illustrated by Fig 6 which displays the sample histograms with the curves of the log-normal densities whose parameters are the estimates computed from the samples.

\begin{figure*}
\centering
\begin{tabular}{ll}
\hspace{-20mm}\includegraphics[scale=0.55]{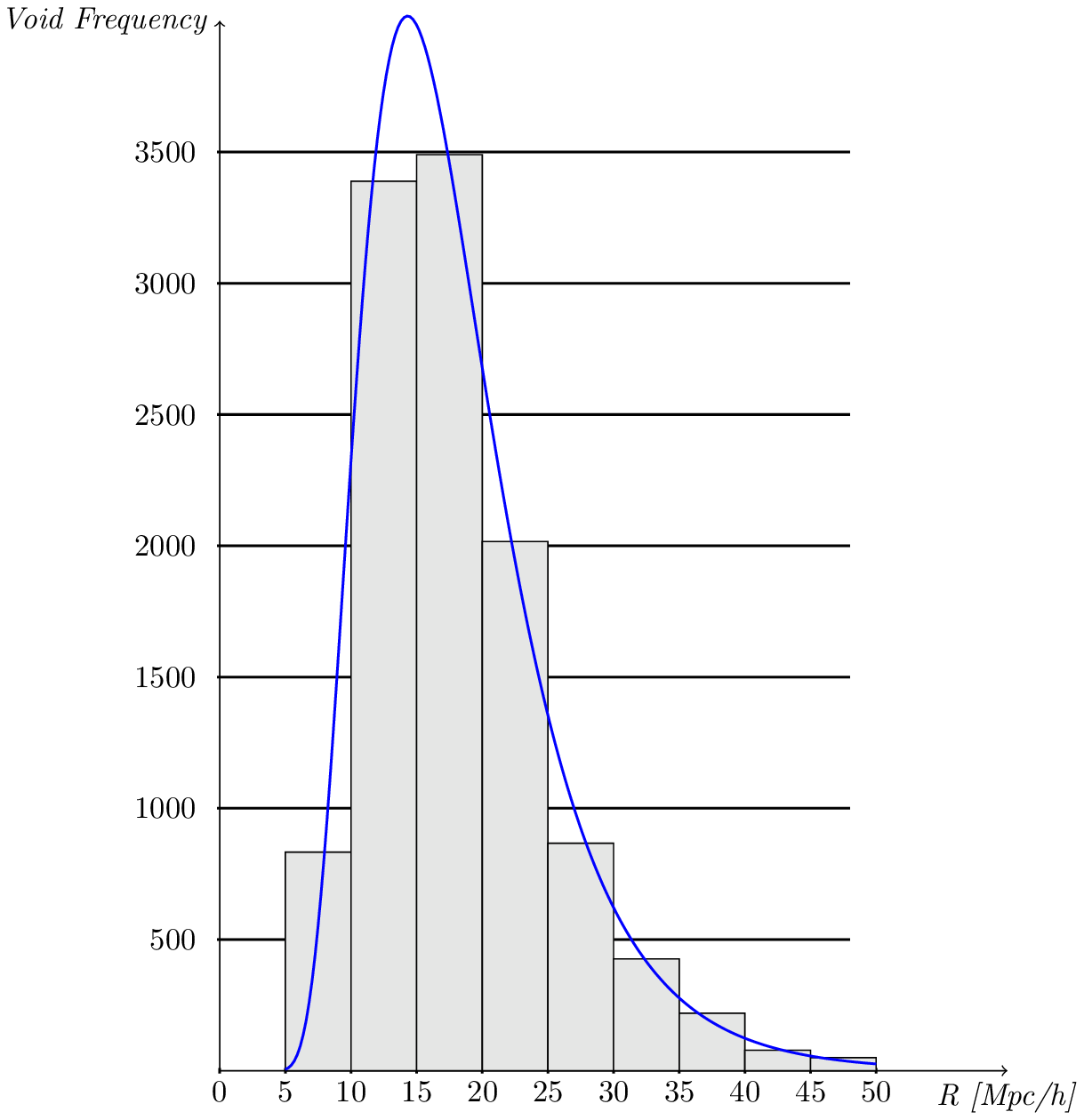}
\hspace{-30mm}\includegraphics[scale=0.55]{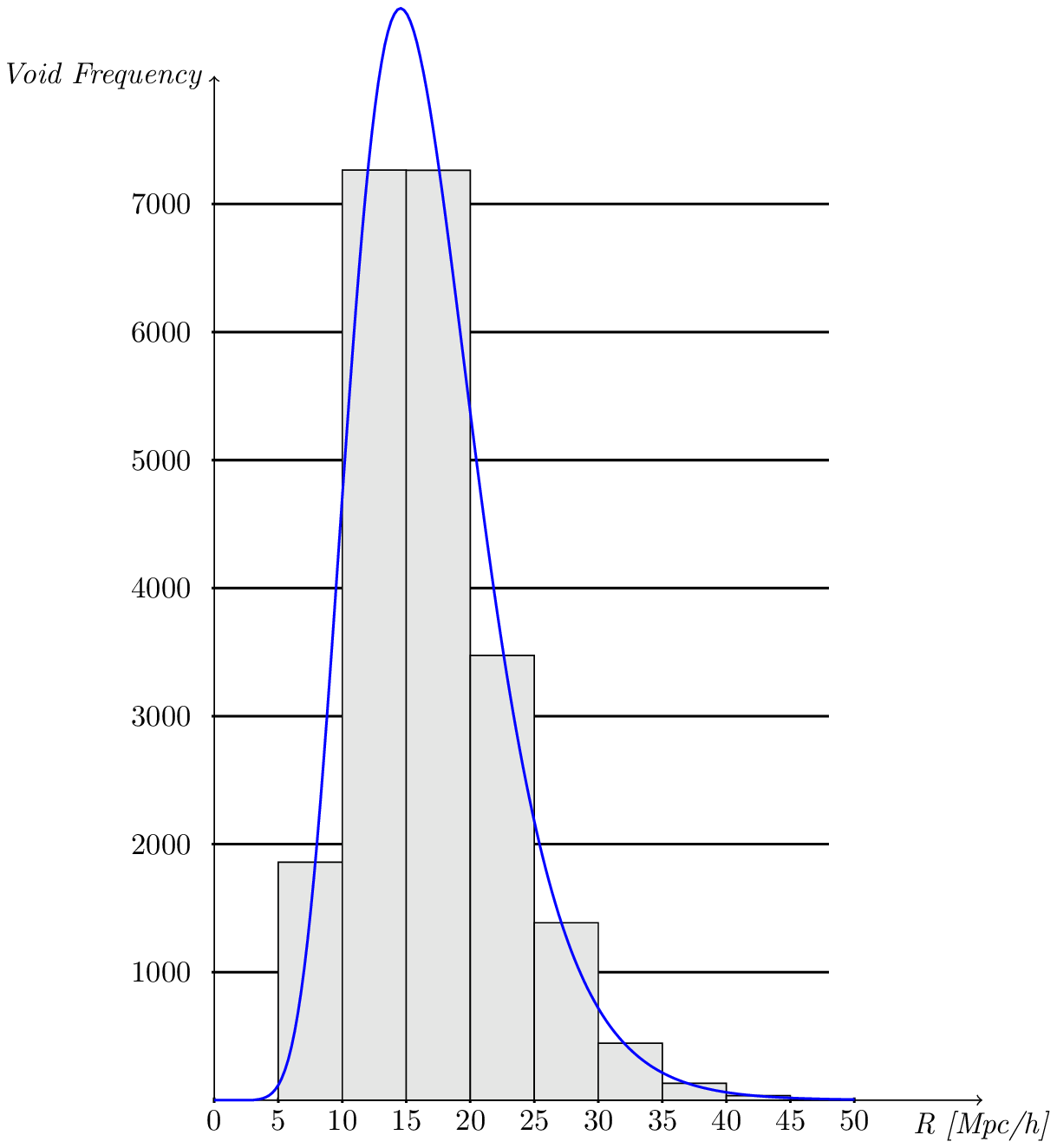}\vspace{-70mm}\\
\hspace{-20mm}\includegraphics[scale=0.5]{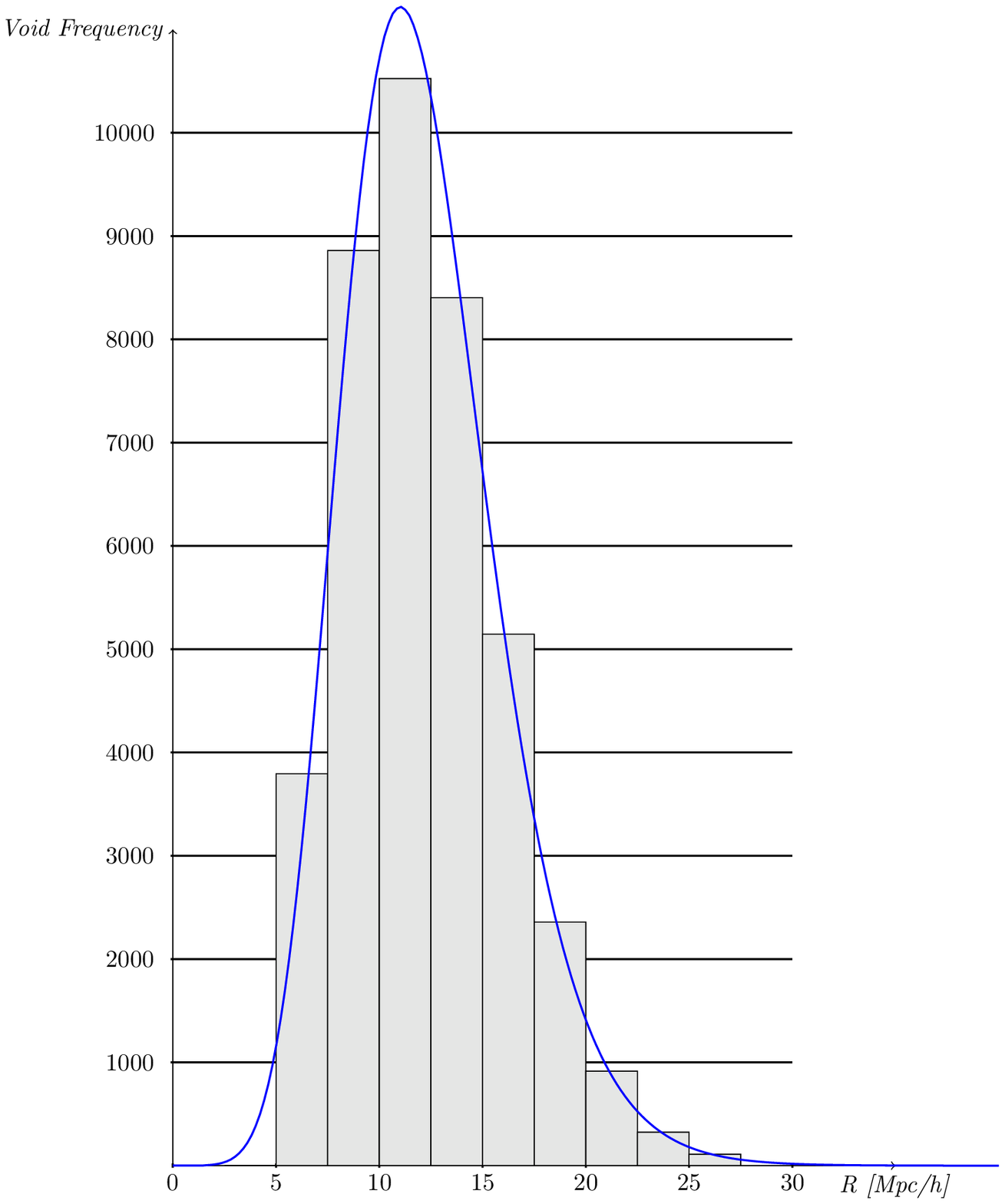}
\end{tabular}
\vspace{-20mm}
\caption{The 3-parameter log-normal void size distributions of Haloes Dense (Upper Left Panel), DM Dense (Upper Right Panel) and DM Full (Lower Right).}
\end{figure*}

\section{Conclusions and Discussion}
Here, extending our previous study of \citep{Pycke2016} to attempt to find a universal void size distribution, we investigate the statistical properties of the void size distribution such as the shape parameters and their relations to the void environment of CVC by using the moment method by following \cite{kotz}. As aforementioned, the moment method is easy to apply. Therefore, we here confirm our previous result on the size distributions of voids which states that the $3$-parameter log-normal distribution gives a satisfactory model of the size distribution of voids, which is obtained from simulation and mock catalogs of CVC; N-body Mock, DM Full, DM Dense, DM Sparse, Haloes Dense, Haloes Sparse, HOD Sparse and HOD Dense (see Fig 6, also Fig 3 in \cite{Pycke2016}).

On the other hand, we should keep in mind that all the data sets of CVC are generated by a single N-body simulation that operates counting scales as NlogN. Therefore the nature of these data sets may enforce us to obtain such a unique void size distribution. At this point being critical is essential before stating that there is a universal void size distribution satisfying the $3$-parameter log-normal. As a result, a thorough investigation of the void size distribution by using other voids catalogs to unveil the truth beyond the relation between the shape of the void size distribution and the void environment has a key importance. Especially, taking into account that \cite{Nadathur2014} point out some problems and inconsistencies in CVC such as the identification of some overdense regions as voids in the galaxy data of the SDSS DR$7$ \citep{Abazajian2009}. Processing from the problems of the CVC, \cite{Nadathur2014} provide an alternative public catalogue of voids, obtained from using an improved version of the same watershed transform algorithm. Therefore, it is essential to extend our analysis of void size distributions to the catalog given by \cite{Nadathur2014}. Again, this is particularly important to confirm whether the $3$-parameter log-normal void size distribution is valid in a different void catalog. If the 3-parameter log-normal distribution fits another simulated/mock void catalogues, then this may indicate that voids have universal (redshift independent) size distributions given by the log-normal probability function.

Apart from this, \cite{Hamaus2014} show that void average density profile can be represented by an empirical function in $\Lambda$CDM N-body simulations by using ZOBOV. This function is universal across void size and redshift. Following this, \cite{Nadathur2015} investigate the density profiles of voids which are identified by again using the ZOBOV in mock luminous red galaxy catalogues from the Jubilee simulation, and in void catalogues constructed from the SDSS LRG and Main Galaxy samples. As a result, \cite{Nadathur2015} show that the scaled density profiles of real voids show a universal behavior over a wide range of galaxy luminosities, number densities and redshifts. Processing from these results, there is a possibility that the $3$-parameter log-normal void size distribution may be a universal distribution for voids in simulated as well as real data samples. That is why, it is critical to extend our analysis to other simulated as well as real data sets.

We also observe that the number of nonzero and zero void central densities in the samples have important effects on the shape of the $3$-parameter log-normal void size distributions. As is seen in Table 1 and Fig 1, if the percentage of number of nonzero central densities reaches $3.84 \% $ in a simulated or mock sample in CVC, then a second population emerges in the void size distribution. This second population presents itself as a second peak in the log-normal size distribution, at larger radius.

Also, we here obtain a linear relation between the maximum tree depth and the skewness of the samples, and this relation is given by equation \ref{eqn:MTD-skew} (see Fig 3). This linear relation indicates that if there is a void in a simulated/mock sample with a high maximum tree depth, then we expect more skewed log-normal distribution. Therefore, there is a direct correlation between the void substructure and the skewness of the void size distribution. The possibility of this relation is mentioned by \cite{Pycke2016}. Therefore, we here confirm that the skewness of a void size distribution is a good indicator of void substructures in a simulated/mock sample. Aforementioned, the minimum radius cut of CVC samples defined by two density-based criterion as given by \cite{sutter2014a} can affect the relation skewness-MTD. The minimum radius cut of CVC is particularly important because it may not only affect the resulting skewness of the data sets but also other shape parameters of the void size distributions, which can violate the confirmation of the log-normal distribution of the samples. That is why, it is important to study raw samples to understand the effect of the minimum radius-cut.

In addition to skewness-MTD linear relation, another linear correlation is obtained between the maximum tree depth and the variance of the sparse samples of CVC (see equation \ref{eqn:MTD-variance-sparse}). As is seen from Fig 4, sparse samples with high maximum tree depth tend to be more dispersed than a sparse sample with lower maximum tree depth. On the other hand, although we obtain a linear relation between the maximum tree depth and the variance of the dense samples (see equation \ref{eqn:MTD-variance-dense}), due to the lack of dense samples with merging tree depth, this relation does not provide enough information to define a relation between these parameters (see Fig 4, red dotted line). But it is obvious that the relation between maximum tree depth and variance shows two different behaviors for the sparse and the dense samples. This is an expected result since variance is the indicator of dispersion by definition. While sparse samples are highly dispersed with high variance values, the dense samples are expected to show lower variance/dispersion than the sparse samples. These relations indicate that there is a direct correlation between the shape parameters of the void size distribution such as skewness, variance and the void substructures. Our next goal is to address the following questions: Is it possible to relate the shape parameters of the void size distribution to the environment in real data samples? Do the shape parameters change in time, indicating the dynamical evolution of the void size distribution? Is the $3$-parameter log-normal void size distribution universal?

\acknowledgments
The authors would like to thank Paul Sutter and his team for constructing and sharing the Cosmic Void Catalog. All void samples used here can be found in folder void catalog: 2014.06.08 at http://www.cosmicvoids.net.


\begin{thebibliography}{}
\bibitem[Abazajian et al. (2009)]{Abazajian2009} Abazajian, K.~N. 2009, \apjs, 182, 543
\bibitem[Bernardeau(1992)]{Bernardeau92} Bernardeau, F. 1992, \apj, 392, 1
\bibitem[Bernardeau(1994)]{Bernardeau94} Bernardeau, F. 1994, \aap, 291, 697
\bibitem[Bouchet et al.(1993)]{Bouchet1993} Bouchet, F.~R., Strauss, M.~A., \& Davis, M. et al. 1993, \apj, 417, 36
\bibitem[Chincarini \& Rood(1975)]{chincarini} Chincarini G., \& Rood H. J. 1975, Nature, 257, 294
\bibitem[Coles \& Jones(1991)]{colesjones91} Coles, P., \& Jones, B. 1991, \mnras, 248, 1
\bibitem[Conover(1980)]{conover80} Conover, W. J. 1980, Practical nonparametric statistics, 2nd Ed., New York, NY., John Wiley \& Sons
\bibitem[Conover(1999)]{conover99} Conover, W. J. 1980, Practical nonparametric statistics, 3rd Ed., New York, NY., John Wiley \& Sons
\bibitem[Croton et al.(2005)]{Croton2005} Croton, D.~J., Farrar, G.~R., \& Norberg, P. et al. 2005, \mnras, 356, 1155
\bibitem[Dawson et al.(2013)]{Dawson2013} Dawson, K.~S. et al. 2013, \aj, 145, 10
\bibitem[Einasto et al (1980)]{einasto} Einasto, J., Joeveer, M., \& Saar E. 1980, Nature, 283, 47
\bibitem[Elizalde \& Gaztanaga(1992)]{elizaldegaztanaga92} Elizalde, E., \& Gaztanaga, E. 1992, \mnras, 254, 247
\bibitem[Fry(1986)]{fry86Voids} Fry, J.~N. 1986, \apj, 306, 358
\bibitem[Goldberg \& Vogeley(2004)]{goldbergvogeley} Goldberg, D.~M., \& Vogeley, M.~S. 2004, \apj, 605, 1
\bibitem[Gregory \& Thompson(1978)]{gregory} Gregory S. A., \& Thompson L. A. 1978, \apj, 222, 784
\bibitem[Hamilton(1985)]{Hamilton1985} Hamilton, A.~J.~S. 1985, \apjl, 292, L35
\bibitem[Hamaus et al.(2014)]{Hamaus2014} Hamaus, N., Sutter, P.~M., \& Wandelt, B.~D. 2014, Physical Review D, 112, 25
\bibitem[Hoyle et al.(2005)]{Hoyle2005} Hoyle, F., Rojas, R.~R., Vogeley, M.~S., \& Brinkmann, J. et al. 2005, \apj, 620, 618
\bibitem[Johnson et al.(1994)]{kotz} Johnson, N.L, Kotz, S, \& Balakrishnan, N. 1994, Continuous Univariate Distributions, Vol. 1 (2nd ed; Wiley)
\bibitem[Kayo et al.(2001)]{Kayo} Kayo, I., Taruya, A., \& Suto, Y. 2001, \apj, 561, 22
\bibitem[Kendall \& Stuart(1977)]{stuart1} Kendall, M., \& Stuart, A. 1977, The advanced theory of statistics. Vol 1: Distribution Theory, 2nd Ed., (New York, NY., Macmillan)
\bibitem[Kofman et al.(1994)]{Kofman} Kofman, L., Bertschinger, E., Gelb, J.~M., Nusser, A., \& Dekel, A. 1994, \apj, 420, 44
\bibitem[Komatsu et al.(2011)]{komatsu2011} Komatsu, E. et al. 2011, \apjs, 192, 18
\bibitem[Lavaux \& Wandelt (2012)]{Lavaux12} Lavaux, G., \& Wandelt, B.~D. 2012, \apj, 754, 109
\bibitem[Manera et al.(2013)]{Manera2013} Manera, M., et al. 2013, \mnras, 428, 1036
\bibitem[Neyrick(2008)]{Neyrinck2008} Neyrinck, M.~C. 2008, \mnras, 386, 2101
\bibitem[Nadathur \& Hotchkiss(2014)]{Nadathur2014} Nadathur, S.,\& Hotchkiss, S. 2014, \mnras, 440, 1248
\bibitem[Nadathur et al. (2015)]{Nadathur2015} Nadathur, S. 2015, \mnras, 449, 3997
\bibitem[Planck Collaboration(2014)]{PlanckCollaboration} Planck Collaboration, 2014, \aap, 571, A19
\bibitem[Russell(2013)]{Esra1}  Russell, E.  2013, \mnras, 436, 3525
\bibitem[Russell(2014)]{Esra2} Russell, E.  2014, \mnras, 438, 1630
\bibitem[Pycke \& Russell (2016)]{Pycke2016} Pycke, J.-R., \& Russell, E.  2016, \apj, 821, 110
\bibitem[Sheth \& van de Weygaert (2004)]{sw} Sheth, R.~K., \& van de Weygaert, R. 2004, \mnras, 350, 517
\bibitem[Sheskin(2011)]{sheskin} Sheskin, D. J. 2011, Handbook of parametric and nonparametric statistical procedures (Boca Raton: FL, CRC Press)
\bibitem[Strauss et al. (2002)]{Strauss2002} Strauss, M.~A. 2002, \aj, 124, 1810
\bibitem[Sutter et al.(2012)]{sutter} Sutter, P.~M., Lavaux, G., Wandelt, B.~D., \& Weinberg, D.~H. 2012, \apj, 761, 44
\bibitem[Sutter et al.(2014a)]{sutter2014a} Sutter, P.~M., Lavaux, G., Hamaus, N., Wandelt, B.~D., Weinberg, D.~H., \& Warren, M.~S., 2014, \mnras, 462, 471
\bibitem[Sutter et al.(2014b)]{sutter2014b} Sutter, P.~M., Lavaux, G., Wandelt, B.~D., Weinberg, D.~H., Warren, M.~S., \& Pisani, A., 2014, \mnras, 442, 3127
\bibitem[Taylor \& Watts (2000)]{TaylorWatts2000} Taylor, A.~N., \& Watts, P.~I.~R. 2000, \mnras, 314, 92
\bibitem[Tinker et al.(2006)]{Tinker2006} Tinker, J.~L., Weinberg, D.~H., \& Zheng, Z., 2006, \mnras, 368, 85
\bibitem[Weygaert \& Platen(2011)]{weygaert11} van de Weygaert R., \& Platen E. 2011, International Journal of Modern Physics Conference Series, 1, 41
\bibitem[White(1979)]{white79} {White}, S.~D.~M. 1979, \mnras, 186, 145
\bibitem[Zehavi et al.(2011)]{Zehavi2011} Zehavi, I. et al. 2011, \apj, 736, 59
\bibitem[Zheng et al.(2007)]{Zheng07} Zheng, Z., Coil, A.~L., \& Zehavi, I. 2007, \apj, 667, 760
\end{thebibliography}
\end{document}